\documentclass[12pt]{article}
\pdfoutput=1
\usepackage{comment}
\usepackage{amsmath}
\usepackage{amssymb}
\usepackage{nccmath}
\usepackage{graphicx}
\usepackage{here}
\usepackage{subcaption}
\usepackage{url}
\usepackage[sort&compress, numbers, merge]{natbib}
\usepackage{braket}
\usepackage{physics}
\usepackage[compat=1.1.0]{tikz-feynhand}
\usepackage{slashed}
\usepackage{mathtools}
\usepackage{bbold}
\usepackage{bbm}
\usepackage{booktabs}
\usepackage[shortcuts]{extdash}
\usepackage[range-phrase={\text{--}}]{siunitx}

\usepackage{tikz}
\usetikzlibrary{shapes,tikzmark,3d}
\usepackage{tikz-3dplot}
\usetikzlibrary{decorations.markings}
\usetikzlibrary{shapes.geometric}
\tikzset{every tikzmarknode/.style={
        draw=red, semithick, inner sep=2pt}
        }
\usepackage{pgfplots}
\pgfplotsset{compat=1.8}
\usepackage{xcolor}
\definecolor{CUDbluegreen}{HTML}{0072B2}    

\definecolor{CUDblue}{RGB}{0,114,178}
\definecolor{CUDorange}{RGB}{230,159,0}
\definecolor{CUDgreen}{RGB}{0,158,115}
\definecolor{CUDpurple}{RGB}{204,121,167}
\definecolor{CUDgray}{RGB}{128,128,128}
\definecolor{CUDnavy}{RGB}{0,45,90}
\definecolor{CUDred}{RGB}{213,94,0}

\newcommand{\alias}[3]{%
  \let#3=#2
  \newcommand{#1}{#3}
  \renewcommand{#2}{{\textcolor{red}{#3}}}
}
\alias{\unw}{\beta}{\origbeta}
\newcommand{\cdA}{D}
\newcommand{\cdF}{D}

\newcommand{\defby}{\coloneqq}
\newcommand{\lagrangian}{\mathcal{L}}

\newcommand{\Vlambda}{\lambda_\phi}
\newcommand{\Hlambda}{\lambda_H}

\alias{\ct}{\dagger}{\origdagger}
\newcommand{\fund}{H}

\newcommand{\U}{\mathrm{U}}
\newcommand{\SU}{\mathrm{SU}}
\newcommand{\pauli}[1]{\tau^{#1}}
\newcommand{\sugen}[1]{\frac{\pauli{#1}}{2}}

\numberwithin{equation}{section}

\newcommand{\version}{arXiv1}
\DeclareRobustCommand{\modifiedat}[2]{%
\ifthenelse{\equal{\version}{#1}}{\textcolor{red}{#2}}{#2}%
}

\usepackage[colorlinks=true, linkcolor=blue, citecolor=blue,
urlcolor=black]{hyperref} 
\usepackage[capitalize]{cleveref}
\crefname{section}{Sec.}{Secs.}
                 
\begin{document}
\def\ps{\mathbf{p}}
\def\PS{\mathbf{P}}
\baselineskip 0.6cm
\def\simgt{\mathrel{\lower2.5pt\vbox{\lineskip=0pt\baselineskip=0pt
           \hbox{$>$}\hbox{$\sim$}}}}
\def\simlt{\mathrel{\lower2.5pt\vbox{\lineskip=0pt\baselineskip=0pt
           \hbox{$<$}\hbox{$\sim$}}}}
\def\simprop{\mathrel{\lower3.0pt\vbox{\lineskip=1.0pt\baselineskip=0pt
             \hbox{$\propto$}\hbox{$\sim$}}}}
\def\tr{\mathop{\rm tr}}
\def\SU{\mathop{\rm SU}}

\def\VH{v_H}
\def\VG{v_\phi}
\def\azimuth{\varphi}
\newcommand{\az}{\azimuth}
\def\zenith{\theta}
\def\higgsprofile{\xi}
\def\Newton{G_N}

\begin{titlepage}

\begin{flushright}
IPMU25-0023
\end{flushright}

\vskip 1.1cm

\begin{center}

{\Large \bf 
Cosmic Strings in Multi-Step Symmetry Breaking
}

\vskip 1.2cm
Akifumi Chitose$^{a}$,
Masahiro Ibe$^{a,b}$, 
Satoshi Shirai$^{b}$ and
Yaxuan Wen$^{a}$
\vskip 0.5cm

{\it

$^a$ {ICRR, The University of Tokyo, Kashiwa, Chiba 277-8582, Japan}

$^b$ {Kavli Institute for the Physics and Mathematics of the Universe
(WPI), \\The University of Tokyo Institutes for Advanced Study, \\ The
University of Tokyo, Kashiwa 277-8583, Japan}

}

\end{center}

\vskip 1.0cm

\abstract{
We investigate cosmic strings arising from a hierarchical gauge symmetry breaking sequence,
$
\mathrm{SU}(2) \times \mathrm{U}(1) \rightarrow \mathrm{U}(1) \times \mathrm{U}(1) \rightarrow \mathrm{U}(1)' \rightarrow \text{Nothing}.
$
This pattern gives rise to two distinct classes of cosmic strings: light, stable strings formed at a later stage, and heavy, metastable strings originating from an earlier stage. Our focus is on the heavy strings, which may decay either before or after the final $\mathrm{U}(1)'$ symmetry is broken.  
We analyze the internal structure of these strings and the magnetic flux sourced by monopole-like configurations that emerge at the endpoints of metastable string segments following their decay.  
Understanding the nature of the magnetic $\mathrm{U}(1)$ fluxes associated with these monopole-like objects is crucial for studying the post-decay evolution of the string network. The post-decay evolution 
influences the resulting gravitational wave signals. 
We show that the magnetic flux carried by string segments can be either confined or unconfined, depending on the specific sequence of symmetry breaking and string decay.

}

\end{titlepage}

\section{Introduction}
Recently, multiple pulsar timing array (PTA) collaborations have reported evidence for a stochastic gravitational‐wave background in the nanohertz band~\cite{NANOGrav:2023gor,EPTA:2023fyk,Reardon:2023gzh,Xu:2023wog}. In light of these observations, cosmic strings have attracted renewed attention as a compelling source of such signals. Cosmic strings are topological defects formed during symmetry-breaking phase transitions like $\mathrm{U}(1) \to \mathrm{Nothing}$ in the early Universe, characterized by nontrivial winding number. Once formed, a cosmic string network evolves and radiates energy predominantly via gravitational waves (see, e.g. Ref.~\cite{Vilenkin:2000jqa}).

Intriguingly, current PTA data favor scenarios with metastable rather than stable cosmic strings. In these models, strings decay at late times, reducing the gravitational-wave spectrum at low frequencies and improving consistency with PTA observations~\cite{Leblond:2009fq,Buchmuller:2019gfy,Buchmuller:2020lbh,Buchmuller:2021mbb,Buchmuller:2023aus}. This has triggered growing interest in the cosmological implications of metastable cosmic strings~\cite{Antusch:2023zjk,Fu:2023mdu,Lazarides:2023rqf,Ahmed:2023rky,Afzal:2023cyp,Maji:2023fhv,Ahmed:2023pjl,Afzal:2023kqs,King:2023wkm,Pallis:2024mip,Maji:2024pll,Antusch:2024nqg,Maji:2024tzg,Maji:2024cwv,Pallis:2024joc,Ahmad:2025dds,Antusch:2025xrs,Hu:2025sxv,Maji:2025thf}.

Metastable cosmic strings commonly emerge in multi-step symmetry breaking sequences of the form $G \to H \to K$, where $\pi_1(H/K) \neq 0$ but $\pi_1(G/K) = 0$~\cite{Vilenkin:1982hm}. 
Here, $\pi_i$ denotes the $i$-th homotopy group. Since $\pi_1(G/K)$ is trivial, the cosmic strings are only metastable and decay via monopole-antimonopole pair production. In fact, the decay is caused by the difference between $\pi_1(H/K)$ and $\pi_1(G/K)$, with which a nontrivial element of $\pi_2(G/H)\simeq \pi_1(H)/\pi_1(G)$ is associated~\cite{Preskill:1992ck}.
As $\pi_2(G/H)$ classifies monopoles emerging at the first symmetry breaking $G\to H$, this indicates the correspondence between the metastable cosmic strings and the monopoles cutting them.
For example, in the simplest realization with $G = \mathrm{SU}(2)$, $H = \mathrm{U}(1)$ and $K = \mathrm{Nothing}$, cosmic strings from $\pi_1(\U(1))=\mathbb{Z}$ are cut by monopoles from $\pi_2(\SU(2)/\U(1))=\mathbb{Z}$.
The structure and decay mechanisms of such metastable strings in minimal symmetry breaking patterns have been investigated in Refs.\,\cite{Vilenkin:1982hm,Preskill:1992ck,Shifman:2002yi,Monin:2008mp,Chitose:2023dam}.

In realistic particle-physics settings, symmetry breaking often proceeds through more intricate chains. For example, in the inflationary model of Ref.\,\cite{Chitose:2024pmz}, which ensures the formation of the long cosmic string network,
accounting for the PTA signal via metastable strings, 
the symmetry breaking proceeds as
\begin{align}
\mathrm{SU}(2) \times \mathrm{U}(1) \to \mathrm{U}(1) \times \mathrm{U}(1) \to \mathrm{U}(1)' \to \mathrm{Nothing}\ .
\end{align}
Such patterns may arise 
in many theoretical frameworks beyond the Standard Model, including grand unified theories and models with extended gauge symmetries. 
Compared to simpler scenarios, the configuration and decay processes of metastable strings arising in extended symmetry breaking patterns are still not well understood. In particular, it is still unclear how the monopoles at the endpoints of string segments, resulting from cosmic string decay, carry magnetic flux in these scenarios.

In this paper, we investigate metastable cosmic strings in this extended breaking pattern as a step toward improving our understanding of such composite topological defects. We demonstrate that the magnetic flux carried by string segments can be either confined or unconfined depending on the order of symmetry breaking and string decay. This flux character affects the subsequent evolution of the segment network and its gravitational wave signals.

The remainder of the paper is organized as follows. In Sec.\,\ref{sec:SetUp}, we review the minimal model $\mathrm{SU}(2)\to\mathrm{U}(1)\to \mathrm{Nothing}$ and the associated topological defects. In Sec.\,\ref{sec:Setup2}, we extend the analysis to include the additional $\mathrm{U}(1)$ symmetry and study the confined/unconfined flux structure. We conclude in the final section.

\section{Review of Metastable String}
\label{sec:SetUp}

\subsection{\texorpdfstring{$
\boldsymbol{\SU(2)}$}{SU(2)} Model }
In this section, we 
review 
the metastable string in a simpler model, where the gauge group is given by a simple SU(2) gauge group which we call  $\SU(2)_\mathrm{M}$.
We assume that the model has an adjoint scalar field $\phi^a$ ($a=1,2,3$)
and a doublet scalar 
field $H_i$ ($i=1,2$) of $\SU(2)_\mathrm{M}$.
We use the following conventions for the covariant derivatives and field strength, 
\begin{align}
    \cdF_\mu \fund&\defby\partial_\mu \fund -  ig \sugen{a}W_\mu^a\fund\ ,\\
    \cdA_\mu \phi^a &\defby \partial_\mu\phi^a + g  \epsilon^{abc}W_\mu^b \phi^c\ , \\
    W_{\mu\nu}^a &\defby \partial_\mu W^a_\nu - \partial_\nu W^a_\mu + g\epsilon^{abc}W^b_\mu W^c_\nu \ , 
\end{align}
where $g $ is the $\SU(2)_\mathrm{M}$ gauge coupling and $\pauli{}$'s are the Pauli matrices and the doublet indices are suppressed.
The Lagrangian of the model is given by
\begin{equation} \label{eq:lagrangian}
\lagrangian=-\frac{1}{2}(\cdA_\mu \phi^a)(\cdA^\mu \phi^a) - (\cdF_\mu \fund)^\ct (\cdF^\mu \fund) - \frac{1}{4}W^a_{\mu\nu}W^{a\mu\nu} -V(\phi,\fund)\ .
\end{equation}
The convention for the Minkowski metric is taken to be $(g_{\mu\nu})=\mathrm{diag}(-1,1,1,1)$.
The scalar potential is given by
\begin{equation}
    V(\phi,\fund)\defby 
    \Vlambda\pqty{\phi^a\phi^a-\VG^2}^2+
    \Hlambda\pqty{\abs{\fund}^2-\VH^2}^2+\lambda_{H\phi} \abs{\pqty{\phi-\frac{\VG}{2}}\fund}^2\ ,
    \label{eq:VphiH}
\end{equation}
where $\phi\defby\phi^a\pauli{a}/2$.
The dimensionless coupling constants $\Hlambda$, $\Vlambda$ 
and $\lambda_{H\phi}$ are taken to be positive.
We also assume that the two breaking scales $\VG$  and $\VH$ are hierarchical, i.e., $\VG \gg \VH (>0)$.
In this paper, we assume that the $W_{\mu\nu}\tilde{W}^{\mu\nu}$ term is vanishing for simplicity. 
See e.g.,  Ref.\,\cite{Chitose:2023bnd}
for the effect of the term on the topological defects (see also Refs.\,\cite{Brummer:2009cs,Long:2014mxa,Hook:2017vyc} for related works).

When $\phi$ takes the trivial configuration,
\begin{equation}
\label{eq:VEV1}
   \langle{\phi^a}\rangle = \VG \delta^{a3}\ ,
\end{equation}
it breaks $\SU(2)_\mathrm{M}$ 
down to $\mathrm{U}(1)$, which we call the $\mathrm{U}(1)_\mathrm{M}$ symmetry in the following.
Due to the $\phi$--$H$ interacting term  with $\lambda_{H\phi}>0$, the second component of $\fund$ obtains a large positive mass squared,  $m_{H_2}^2 = \lambda_{H\phi }v_\phi^2$,  and hence, 
the VEV of $H$ is aligned to,
\begin{align}
    \langle H\rangle =  
    \mqty(\VH\\0)\ ,
\end{align}
which breaks the remaining $\mathrm{U}(1)_\mathrm{M}$ symmetry.
In this way, the model achieves a successive symmetry breaking,
\begin{align}
    {\SU}(2)_\mathrm{M} \xrightarrow{\expval{\phi}} \mathrm{U}(1)_\mathrm{M} 
    \xrightarrow{\expval{H}} 
    \mathrm{Nothing}\ .
\end{align}

Finally, we note that the scalar potential in Eq.\,\eqref{eq:VphiH} has a special structure, where $\phi$ and $H$ are decoupled except for the $\lambda_{H\phi}$ term. We adopt this structure for simplicity. Such a potential can be achieved in supersymmetric models, for example by imposing a discrete $R$-symmetry~\cite{Chitose:2024pmz}. However, we will not pursue those details here.

\subsection{Monopoles}
Let us first consider 
the 't~Hooft-Polyakov monopole~\cite{tHooft:1974kcl,Polyakov:1974ek},
which appears at the first phase transition, 
$\SU(2)_\mathrm{M}\to\mathrm{U}(1)_\mathrm{M}$.
In the following, 
we discuss only the asymptotic behavior of the monopole solution.
The asymptotic behavior of the
static monopole configuration
with a unit winding number is given by
\begin{align}
\label{eq:heg}
\phi^a\to \VG \displaystyle{\frac{x^a}{r}}\ , \quad
W_{0}^a = 0 \ ,\quad
g  W^{a}_{i}\to
\displaystyle{\frac{\epsilon^{aij}x^j}{r^2}}\ ,\quad (i,j=1,2,3)\ ,
\end{align}
where $(x^0,x^1,x^2,x^3)$
are the spacetime coordinates and 
$r\defby\sqrt{x_1^2+x_2^2+x_3^2}$.
The monopole solution asymptotes to the above behavior for $r \gg (g \VG)^{-1}$.
As we assume the hierarchical breaking,
i.e.,  $\VG \gg \VH$, 
we neglect the effect of $\fund$, and set it to zero.
Notice that the $\SU(2)_\mathrm{M}$ gauge symmetry is restored at the origin of the monopole.

To see the magnetic field around the monopole,
it is convenient to define the effective $\mathrm{U}(1)_\mathrm{M}$
field strength as
\begin{equation}
\label{eq:effectiveF}
W_{\mathrm{M}\mu\nu} := \frac{1}{\VG}\phi^a W^a_{\mu\nu}\ ,
\end{equation}
(see e.g., Ref.\,\cite{Shifman:2012zz}).
The only non-vanishing components of $W_{\mathrm{M}\mu\nu}$ are 
\begin{align}
    g W_{\mathrm{M}}{}^{i j } \to -
\frac{\epsilon^{ijk}x^k}{r^3} \ , \quad (i,j=1,2,3)\ .
\end{align}
Hence, the magnetic charge of the monopole is given by
\begin{equation}
\label{eq: monopole flux}
    Q^m:=  \int_{r \to \infty} 
    \dd
S_{i} ~ g  \frac{\varepsilon^{ijk}}{2} W_{\mathrm{M}jk} =  -4\pi \ ,
\end{equation}
where $\dd{S}_{i}$ is the surface element of a two dimensional sphere surrounding the monopole.

\subsection{Cosmic Strings}
\label{sec:CosmicString}
Let us now discuss the Abrikosov–Nielsen–Olesen (ANO) string~\cite{Abrikosov:1956sx,Nielsen:1973cs} that forms at the second stage of symmetry breaking.
Here, we assume the trivial $\phi$ configuration given in Eq.\,\eqref{eq:VEV1}. 
In the limit $\VG \to \infty$, the fields $W^{1,2}$, $\phi$, and $H_2$ become heavy and only  $W^3$ and $H_1$
are left in the low-energy effective theory. The low-energy effective theory can thus be treated as a $\U(1)_\mathrm{M}$ gauge theory with a single complex scalar field $H_1$ of charge $1/2$. 
The covariant derivative is
\begin{align}
D_\mu H_1 = \left(\partial_\mu -  \frac{i}{2}g W_\mu^3\right) H_1\ .
\end{align}

The asymptotic behavior of the static string solution aligned along the $x^3$-axis is given by (see e.g., Ref.\,\cite{Vilenkin:2000jqa})
\begin{align}
\label{eq:string ansatz1}
H_1 &\to\VH e^{i n\azimuth} \ , \\
\label{eq:string ansatz2}
g  W^3_i &\to - {2n}\frac{\epsilon_{ij}x^j}{\rho^2}\ ,~~~~(i,j=1,2)\ , \\
W^3_{0} &= W^3_{3} = 0 \ ,
\end{align}
where $n \in \mathbbm{Z}$ is the winding number of the string. We adopt cylindrical coordinates with $x_1 = \rho\cos\varphi$
and $x_2 = \rho \sin\varphi$.
Note again that the second component $H_2$ acquires a large mass and is therefore set to zero in the low-energy description.
The cosmic string solution asymptotes to the above configuration exponentially for $\rho \gg (g v_H)^{-1}$. In this regime, $D_\mu H_1 \to 0$ exponentially as well, ensuring finite string tension.

The winding number is related to the magnetic flux along the string by
\begin{align}
\label{eq: string flux}
\int \dd[2]{x}\, g W^3_{12}
= \oint_{\rho\to\infty} g W^3_i \dd{x}^i
= 4\pi n \ .
\end{align}
Thus, for $|n| = 1$, the magnetic flux carried by the string matches the magnetic charge of a magnetic monopole or antimonopole.
\subsection{Monopoles Connected by Cosmic String}
\label{sec:monopole connection}
Thus far, we have discussed the magnetic monopole in SU(2) breaking and the cosmic string in U(1) breaking. However, 
in the full theory, the gauge symmetry is completely broken,
\begin{align}
    \SU(2)_\mathrm{M}\,    {\xrightarrow{\expval{\phi},\,\expval{H}}}\,
    \text{Nothing}\ .
\end{align}
Since $\pi_2(\SU(2)) = \pi_1(\SU(2)) = 0$, no 
monopoles
and cosmic strings exist.
In the following, we discuss how monopoles and strings
once appeared at the each step of symmetry breaking disappear.

To see the fate of a monopole, it is useful
to adopt the gauge consisting of  
two slightly overlapping charts covering the northern and southern hemispheres,
\begin{align}
\label{eq:NSchart}
    U_{N} &= 
    \left\{(r, \zenith,\azimuth)|0 \le \zenith \le {\pi}/{2} + \varepsilon
    ,\, r>R\right\}\ ,\\
    U_{S} &= 
    \left\{(r, \zenith,\azimuth)|{\pi}/{2}-\varepsilon \le \zenith \le \pi
    ,\, r>R\right\}\ .
\end{align}
Here, $\zenith$ is the 
zenith angle, $\varepsilon$ is a small positive number, and $R$ is 
some length scale satisfying 
$R\gg (g \VG)^{-1}$.
Then, we transform the monopole configuration in Eq.\,\eqref{eq:heg} as
\begin{align}
    \phi^a \sugen{a} &\to  \phi{}_{N,S}^a \sugen{a} = g_{N,S} \phi^a \sugen{a} g_{N,S}^\ct\ , \\
W^a_{i}\sugen{a} &\to  W_{N,S\,i}^a \sugen{a} = g_{N,S} W_{i}^a\sugen{a} g_{N,S}^\ct +
\frac{i}{g } g_{N,S} \partial_ig_{N,S}^\ct\ ,
\label{eq:localDP}
\end{align}
with
\begin{align}
\label{eq:combedgauge}
    &g_N =\left(
    \begin{array}{cc}
      c_{\zenith/2}  &  e^{-i\azimuth}s_{\zenith/2} \\
      - e^{i\azimuth}s_{\zenith/2}& c_{\zenith/2}
    \end{array}
    \right)\ ,
    ~~~~~~~ g_S =\left(
    \begin{array}{cc}
       e^{i\azimuth}c_{\zenith/2}  & s_{\zenith/2} \\
      -s_{\zenith/2}& e^{-i\azimuth}c_{\zenith/2}
    \end{array}
    \right)
\end{align}
in each chart.
We call this the combed gauge.

In the combed gauge, the asymptotic behavior of the monopole at $r \gg (g\VG)^{-1}$ is given by
\begin{align}
\label{eq:phiN}
    \phi{}_N^a &\to \VG \delta^{a3}\ ,\qquad
    g W^{a}_{N} \to \delta^{a3} (\cos\zenith -1) \dd{\azimuth}
\end{align}
in the $U_N$ chart and
\begin{align}
\label{eq:phiS}
    \phi{}^a_S  &\to \VG \delta^{a3}\ ,\qquad
     g W^{a}_{S}\to \delta^{a3} 
    (\cos\zenith+1)  \dd{\azimuth}
\end{align}
in the $U_S$ chart. 
Here, we denote the gauge potential as a one-form field.
The other components
$W^{a}_{N,S}$ ($a=1,2$) vanish asymptotically for $r \gg (g \VG)^{-1}$.

In the combed gauge, the VEV of the adjoint scalar takes the same form as the vacuum in Eq.\,\eqref{eq:VEV1}, and thus the $\U(1)_\mathrm{M}$ gauge potential corresponds to $W_\mu^3$ far from the monopole. However, due to the monopole configuration, $W^3_{N,S}$ are related at the equator $\zenith \sim \pi/2$ via the non-trivial transition function
\begin{align}
\label{eq:transition}
    t_{NS} = e^{2i\azimuth}\ ,
\end{align}
under which
\begin{align}
    g W^3_S = g W^3_N + 2\,\dd{\azimuth}\ .
\end{align}

Now we consider the behavior of the scalar field $H_1$ around the monopole. Suppose that $H_1$ takes a constant VEV $\VH$ in the northern hemisphere far from the monopole, i.e., for $r \gg (g \VG)^{-1}$. The $\U(1)_\mathrm{M}$ magnetic flux of $W^3$ is then expelled from the northern hemisphere by the Meissner effect, so that
\begin{align}
    W^3_N = 0
\end{align}
in this region.

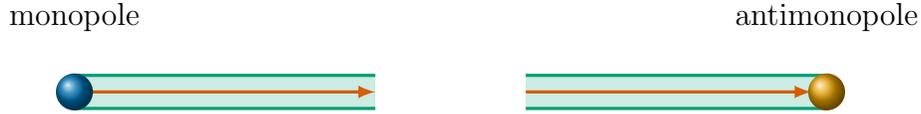
\begin{figure}[t]
    \centering
\begin{tikzpicture}[scale=2]
  \begin{scope}
    \fill[CUDgreen, opacity=0.2] (0,-0.11) rectangle (2,0.11);
    \draw[-latex, very thick, CUDred] plot [smooth] coordinates {(0,0) (2,0)};
    \draw[-,      very thick, CUDgreen, yshift=-0.11cm] plot [smooth] coordinates {(0,0) (2,0)};
    \draw[-,      very thick, CUDgreen, yshift= 0.11cm] plot [smooth] coordinates {(0,0) (2,0)};
    \shade[ball color=CUDblue] (0,0) circle (3.5pt);
    \node at (0,0.5) {monopole};
  \end{scope}

  \begin{scope}[xshift=3cm]
    \fill[CUDgreen, opacity=0.2] (0,-0.11) rectangle (2,0.11);
    \draw[-latex, very thick, CUDred] plot [smooth] coordinates {(0,0) (1.9,0)};

    \draw[-,      very thick, CUDgreen, yshift=-0.11cm] plot [smooth] coordinates {(0,0) (2,0)};
    \draw[-,      very thick, CUDgreen, yshift= 0.11cm] plot [smooth] coordinates {(0,0) (2,0)};

    \shade[ball color=CUDorange] (2,0) circle (3.5pt);

    \node at (2,0.5) {antimonopole};
  \end{scope}
\end{tikzpicture}
    \caption{(Anti)monopole attached to a cosmic string. The magnetic flux of the (anti)monopole is confined within the string.}
    \label{fig:attached}
\end{figure}

In the southern hemisphere below the overlap, the scalar and gauge fields take the form
\begin{align}
\label{eq:phiS2}
    H_{1\,S} &= e^{i\azimuth} H_{1\,N} \,, \\
    g W^3_S &= 2\,\dd{\azimuth} \,,
\end{align}
due to the transition function \eqref{eq:transition}. Thus, the trivial configuration of $H_1$ in the northern hemisphere induces a winding configuration with $n = 1$ in the southern hemisphere, which is a cosmic string attached to a monopole.

Importantly, the magnetic charge of the monopole matches the magnetic flux confined inside the cosmic string (see Eqs.\,\eqref{eq: monopole flux} and \eqref{eq: string flux}). In the $\U(1)_\mathrm{M}$-broken phase, this flux cannot spread out into the vacuum nor terminate arbitrarily, as required by the Bianchi identity. Consequently, each monopole must be attached to a cosmic string carrying its flux (see Fig.\,\ref{fig:attached}). Antimonopoles similarly serve as endpoints of the string. The resulting monopole–antimonopole pair is connected by a string segment and ultimately annihilates.

\subsection{String Decay}

As discussed in Sec.\,\ref{sec:CosmicString}, in regions without monopoles, the breaking $\U(1)_\mathrm{M} \to \text{Nothing}$ can lead to the formation of cosmic strings. 
For instance, if the $\SU(2)_\mathrm{M}$ breaking occurs before inflation and monopoles are inflated away, long cosmic strings may form.

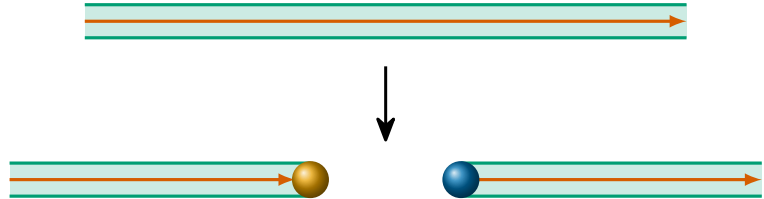
\begin{figure}
    \centering

\begin{tikzpicture}[scale=2]

  \fill[CUDgreen, opacity=0.2] (0,-0.11) rectangle (4,0.11);
  \begin{scope}

    \draw[-latex, very thick, CUDred] plot [smooth] coordinates {(0,0) (4,0)};

    \draw[-,      very thick, CUDgreen, yshift=-0.11cm] plot [smooth] coordinates {(0,0) (4,0)};
    \draw[-,      very thick, CUDgreen, yshift= 0.11cm] plot [smooth] coordinates {(0,0) (4,0)};
  \end{scope}

  \draw[line width=0.4mm, -{Stealth[length=4mm,round]}]
    (2,-0.3) -- (2,-0.8);
\end{tikzpicture}

\vspace{.5em}

\begin{tikzpicture}[scale=2]

  \begin{scope}
    \fill[CUDgreen, opacity=0.2] (0,-0.11) rectangle (2,0.11);
    \draw[-latex, very thick, CUDred] plot [smooth] coordinates {(0,0) (1.9,0)};
    \draw[-,      very thick, CUDgreen,   yshift=-0.11cm] plot [smooth] coordinates {(0,0) (2,0)};
    \draw[-,      very thick, CUDgreen,   yshift= 0.11cm] plot [smooth] coordinates {(0,0) (2,0)};
    \shade[ball color=CUDorange] (2,0) circle (3.5pt);
  \end{scope}

  \begin{scope}[xshift=3cm]
    \fill[CUDgreen, opacity=0.2] (0,-0.11) rectangle (2,0.11);
    \draw[-latex, very thick, CUDred] plot [smooth] coordinates {(0,0) (2,0)};
    \draw[-,      very thick, CUDgreen,   yshift=-0.11cm] plot [smooth] coordinates {(0,0) (2,0)};
    \draw[-,      very thick, CUDgreen,   yshift= 0.11cm] plot [smooth] coordinates {(0,0) (2,0)};
    \shade[ball color=CUDblue] (0,0) circle (3.5pt);
  \end{scope}
\end{tikzpicture}
    \caption{Decay of a cosmic string via monopole–antimonopole pair nucleation. The magnetic flux, represented by red arrows, is confined within the string.}
    \label{fig:string breaking}
\end{figure}

These strings are metastable, that is, a monopole–antimonopole pair can nucleate inside the cosmic string, leading to its spontaneous decay (see Fig.\,\ref{fig:string breaking}). This is a quantum tunneling process: while the string is classically stable, it is quantum mechanically unstable.
In the limit where the width
of the string and the size
of the monopole are negligible, the decay rate per unit length is given by~\cite{Preskill:1992ck}%
\footnote{This approximation is valid only when $\sqrt{\kappa} \gg 10$--$20$~\cite{Chitose:2023dam}.}
\begin{align}
    \Gamma_d = \frac{\mu_{\rm str}}{2\pi}\,e^{-\pi\kappa}\  ,\quad
    \sqrt{\kappa} \simeq \frac{M_M}{\sqrt{\mu_{\rm str}}} \ .
    \label{eq:rootkappaPreskill}
\end{align}
Here, $M_M$ and $\mu_\mathrm{str}$ are the monopole mass and the string tension, respectively:
\begin{align}
    M_M \sim \frac{4\pi v_\phi}{g} \ , \quad 
    \mu_\mathrm{str} \sim 2\pi v_H^2 \ ,
\end{align}
(see also Ref.\,\cite{Chitose:2023dam} for
numerical coefficients). 
Thus, the string lifetime strongly depends on $\kappa$, and becomes particularly long in the regime $\kappa \gg 1$.

\section{Cosmic Strings in \texorpdfstring{$\boldsymbol{\SU(2)_\mathrm{M}\times \mathrm{U(1)}_\mathrm{P}\to \mathrm{Nothing}}$}{SU(2)M→U(1)P→Nothing}}
\label{sec:Setup2}
In this section, we consider a more elaborate symmetry-breaking pattern involving an additional $\mathrm{U}(1)$ gauge symmetry, U(1)$_\mathrm{P}$. The sequence of the symmetry breaking is then assumed to be
\begin{align}
    \SU(2)_\mathrm{M} \times \mathrm{U}(1)_\mathrm{P} \to \mathrm{U}(1)_\mathrm{M} \times \mathrm{U}(1)_\mathrm{P} \to 
    \mathrm{U}(1)_\mathrm{X}
    \to \mathrm{Nothing}\ .
\end{align}
We will show that, depending on whether string decay occurs before or after the breaking of the $\mathrm{U}(1)_\mathrm{X}$ symmetry, the resulting string segments carry either confined or unconfined magnetic flux.

The model studied in this work is based on an $\mathrm{SU}(2)\times \mathrm{U}(1)$ gauge symmetry, which shares the same gauge group structure with the electroweak string~\cite{Nambu:1977ag, Manton:1983nd, Klinkhamer:1984di, Vachaspati:1992jk}. 
In the electroweak theory, only the 
doublet Higgs obtains VEV to break $\SU(2)\times\U(1)$ to $\U(1)$. 
In that case, the
cosmic strings can be classically stable only when the $\mathrm{SU}(2)$ gauge coupling is much smaller than that of the $\mathrm{U}(1)$ coupling~\cite{James:1992zp, James:1992wb}, placing the system near the semilocal string regime~\cite{Vachaspati:1991dz}.

In contrast, our setup includes an $\mathrm{SU}(2)$ triplet scalar field that breaks the $\mathrm{SU}(2)$ symmetry at a high energy scale. 
In this setup, the lifetime of cosmic strings is governed by the ratio of the monopole mass to the string tension (see Eq.\,\eqref{eq:rootkappaPreskill}).
Thus, unlike electroweak strings, our model can accommodate long‐lived metastable strings even when the SU(2)$_\mathrm{M}$ gauge coupling is not weak.

\subsection{\texorpdfstring{$
\boldsymbol{\SU(2)_\mathrm{M}\times \mathrm{U}(1)_\mathrm{P}}$}{SU(2)M×U(1)P} Model}
\begin{table}[t]
\centering
\renewcommand{\arraystretch}{1.2}
\caption{Gauge charges of the fields under $\mathrm{SU}(2)_\mathrm{M} \times \mathrm{U}(1)_\mathrm{P}$.
We also show the mass eigenstate of the gauge boson $Z_\mu$ and $X_\mu$ (see Eq.\,\eqref{eq:mixing}).
See main text for the definition of symbols.}
\begin{tabular}{@{}cccl@{}}
\toprule
 & $\mathrm{SU}(2)_\mathrm{M}$ & $\mathrm{U}(1)_\mathrm{P}$  & Remarks \\
\midrule
$\phi^a$ & $\mathbf{3}$ & 0 & $\expval{\phi^3}$ breaks $\mathrm{SU}(2)_\mathrm{M} \to \mathrm{U}(1)_\mathrm{M}$ \\
$H = \begin{pmatrix} H_1 \\ H_2 \end{pmatrix}$ & $\mathbf{2}$ & $\displaystyle{\frac{1}{2}}$ & $\expval{H_1}$ breaks $\mathrm{U}(1)_\mathrm{M} \times \mathrm{U}(1)_\mathrm{P} \to \mathrm{U}(1)_\mathrm{X}$ \\
$\eta$ & $\mathbf{1}$ & $Q_\eta$ & $\expval{\eta}$ breaks $\mathrm{U}(1)_\mathrm{X} \to \text{Nothing}$ \\
$W^a_\mu$ & $\mathbf{3}$ & 0 & $\mathrm{SU}(2)_\mathrm{M}$ gauge bosons \\
$A_\mu$ & $\mathbf{1}$ & – & $\mathrm{U}(1)_\mathrm{P}$ gauge boson \\
\cmidrule{1-4}
$Z_\mu$ & – & – & $m_Z^2 \simeq g_Z^2v_H^2/2$ \\
$X_\mu$ & – & – & $m_X^2 \simeq 2g_X^2 Q_\eta^2v_\eta^2$ \\
\bottomrule
\end{tabular}
\label{tab:fields}
\end{table}

We extend the $\SU(2)_\mathrm{M}$ model of Sec.\,\ref{sec:SetUp} by introducing an additional gauge symmetry, $\mathrm{U}(1)_\mathrm{P}$. We assume that the doublet field $H$ in the previous model also carries a $\mathrm{U}(1)_\mathrm{P}$ charge, $Q_H = 1/2$. Furthermore, we introduce a new $\SU(2)_\mathrm{M}$ singlet scalar field $\eta$, which has $\mathrm{U}(1)_\mathrm{P}$ charge $Q_\eta$. The adjoint scalar $\phi^a$ remains neutral under $\mathrm{U}(1)_\mathrm{P}$.
In Table\,\ref{tab:fields}, we show the list of fields in this model. 

With this extension, the covariant derivatives of the doublet scalar $H$ and the singlet scalar $\eta$ are given by
\begin{align}
    D_\mu H &:= \partial_\mu H - i g \frac{\tau^a}{2} W_\mu^a H - \frac{i}{2} g' A_\mu H\ , \\ 
    D_\mu \eta &:= \partial_\mu \eta - i g' Q_\eta A_\mu \eta\ .
\end{align}
Here, $A_\mu$ and $g'$ denote the gauge field and gauge coupling constant of $\mathrm{U}(1)_\mathrm{P}$, respectively. 
The normalization of $g'$ is chosen so that the $\mathrm{U}(1)_\mathrm{P}$ gauge charge of $H$ is fixed to $1/2$.
As in the previous model, we assume that the scalar potential for $\phi^a$ and $H$ is given by Eq.\,\eqref{eq:VphiH}, and that they acquire VEVs $v_\phi$ and $v_H$. We further assume that $\eta$ develops a VEV $v_\eta$, satisfying the hierarchy
\begin{align}
    v_\phi \gg v_H \gg v_\eta\ .
\end{align}

In this setup, we focus on the low-energy theory with $\mathrm{U}(1)_\mathrm{M} \times \mathrm{U}(1)_\mathrm{P}$ gauge symmetry. The scalar potential for $H_1$ and $\eta$ is assumed to take the form
\begin{align}
    V(H,\eta) = \lambda_H \left(|H_1|^2 - v_H^2\right)^2 + \lambda_\eta \left(|\eta|^2 - v_\eta^2\right)^2\ ,
    \label{eq:VH_1eta}
\end{align}
with $\lambda_H, \lambda_\eta > 0$. The decoupling of $H_1$ and $\eta$ as in Eq.\,\eqref{eq:VH_1eta} can be realized in supersymmetric models~\cite{Chitose:2024pmz}. 

Let us discuss the trivial vacuum in the extended model. The initial $\SU(2)_\mathrm{M}$ breaking triggered by the VEV of $\phi$ remains unchanged and leads to the vacuum configuration given in Eq.\,\eqref{eq:VEV1}, leaving $\mathrm{U}(1)_\mathrm{M}$ unbroken. 
At the next stage, 
$H_1$ develops a VEV, whereas $\eta$ remains without one.
Then, the VEV of $H_1$ 
induces a mixing between $W^3_\mu$ and $A_\mu$ through the kinetic term of $H$,
\begin{align}
    \mathcal{L} \supset 
    \left| \frac{-i}{2} 
    \begin{pmatrix}
        g W^3_\mu + g' A_\mu & 0 \\
        0 & -g W^3_\mu + g' A_\mu
    \end{pmatrix}
    \begin{pmatrix}
        v_H \\ 0
    \end{pmatrix} \right|^2
    \  .
\end{align}
By diagonalizing the mass matrix of $W^3_\mu$ and $A_\mu$,
we obtain
a massive $Z$ boson and a massless $X$ boson,
\begin{align}
\label{eq:mixing}
    \begin{pmatrix}
        Z_\mu \\
        X_\mu
    \end{pmatrix}
    =
    \begin{pmatrix}
        c_\alpha & s_\alpha \\
        -s_\alpha & c_\alpha
    \end{pmatrix}
    \begin{pmatrix}
        W^3_\mu \\
        A_\mu
    \end{pmatrix}\ , \quad
    c_\alpha := \frac{g}{\sqrt{g^2 + g'^2}}\ , \quad
    s_\alpha := \frac{g'}{\sqrt{g^2 + g'^2}}\ .
\end{align}
The $Z$ boson mass and the mass of the radial component of $H_1$ are given by
\begin{align}
    m_Z^2 &= \frac{1}{2}(g^2 + g'^2) v_H^2\ , \\
    m_H^2 &= 4 \lambda_H v_H^2\ .
\end{align}

The remaining  $\mathrm{U}(1)_\mathrm{X}$ symmetry
is subsequently broken 
 by the VEV of $\eta$ at the final stage of symmetry breaking.
This gives rise to masses for the $X_\mu$ gauge boson and the radial component of $\eta$,
\begin{align}
    m_X^2 &= 2 g_X^2 Q_\eta^2 v_\eta^2\ , \\
    m_\eta^2 &= 4 \lambda_\eta v_\eta^2\ ,
\end{align}
where the effective gauge coupling of $\mathrm{U}(1)_\mathrm{X}$ is
\begin{align}
    g_X = \frac{g g'}{\sqrt{g^2 + g'^2}}\ .
\end{align}
These expressions are valid to leading order in the expansion in $v_\eta^2/v_H^2$ (see Appendix~\ref{sec:mixing}).

Putting everything together, the symmetry breaking sequence in the extended model is
\begin{align}
    \SU(2)_\mathrm{M} \times \mathrm{U}(1)_\mathrm{P}
    \xrightarrow{\langle\phi\rangle} \mathrm{U}(1)_\mathrm{M} \times \mathrm{U}(1)_\mathrm{P}
    \xrightarrow{\langle H\rangle} \mathrm{U}(1)_\mathrm{X}
    \xrightarrow{\langle\eta\rangle} \mathrm{Nothing}\ .
\end{align}
As in the previous model, magnetic monopoles appear during the first symmetry breaking step. Metastable strings appear at the next stage, i.e.,
$\mathrm{U}(1)_\mathrm{M}\times \mathrm{U}(1)_\mathrm{P} \to \mathrm{U}(1)_\mathrm{X}$.
In addition, the final symmetry breaking also produces topologically stable cosmic strings, since $\pi_1[\SU(2)_\mathrm{M} \times \mathrm{U}(1)_\mathrm{P}] = \mathbbm{Z}$ remains nontrivial in the full theory.

As mentioned earlier, the structure of the magnetic flux surrounding a broken string segment plays a crucial role in determining the resulting phenomenology---for example, how the string’s energy is transferred to the thermal plasma.
In the following, we examine the magnetic flux structure in the extended model by considering two distinct cases, classified by the order of the decay time of the metastable string, $t_d$,%
\footnote{In terms of the decay width per unit length, $\Gamma_d$, the decay time for a long string of length $\ell$ (assumed to be of order the Hubble length) is approximately given by $t_d \simeq (\Gamma_d \ell)^{-1}$.}
and the time of $\mathrm{U}(1)_\mathrm{X}$ symmetry breaking, $t_\mathrm{X}$,
\begin{enumerate}
    \item[A.] The string decays well before $\mathrm{U}(1)_\mathrm{X}$ breaking, i.e., $t_\mathrm{X} \gg t_d$.
    \item[B.] The string decays well after $\mathrm{U}(1)_\mathrm{X}$ breaking, i.e., $t_\mathrm{X} \ll t_d$.
\end{enumerate}

\subsection{
Magnetic Flux Along 
\texorpdfstring{$\boldsymbol{Z}$}{Z}-string and \texorpdfstring{$\boldsymbol{X}$}{X}-string}
Before analyzing the more intricate situation where cosmic strings from the second and the third stages overlap, let us first examine the cosmic strings formed at each stage separately.
We call the cosmic string appearing at 
$\mathrm{U}(1)_\mathrm{M}\times \mathrm{U}(1)_\mathrm{P} \to \mathrm{U}(1)_\mathrm{X}$ the $Z$-string,
and the one appearing at $\mathrm{U}(1)_\mathrm{X}\to \mathrm{Nothing}
$ the $X$-string.
The $Z$-string corresponds to the metastable string discussed in the previous section, while the $X$-string is a stable string.

\begin{figure}
    \centering
\begin{tikzpicture}[scale=1.4]

  \begin{scope}

    \fill[CUDgreen, opacity=0.2] (0,-0.2) rectangle (4,0.2);

    \draw[-latex, very thick, CUDnavy, yshift=-0.08cm] plot [smooth] coordinates {(0,0) (4,0)};

    \draw[-,      very thick, CUDgreen, yshift=-0.2cm] plot [smooth] coordinates  {(0,0) (4,0)};
    \draw[-,      very thick, CUDgreen, yshift= 0.2cm] plot [smooth] coordinates  {(0,0) (4,0)};

    \draw[-latex, very thick, CUDred, yshift=0.08cm] plot [smooth] coordinates {(0,0) (4,0)};

    \node[anchor=east] at (10,0)
      {$Z$-string (red:\,$W^3$-flux, navy:\,$A$-flux)};
  \end{scope}

  \begin{scope}[yshift=-1cm]

    \fill[CUDgray, opacity=0.09] (0,-0.2) rectangle (4,0.2);

    \draw[latex-, very thick, CUDnavy, yshift=-0.08cm] plot [smooth] coordinates {(0,0) (4,0)};

    \draw[-,      very thick, CUDgray,   yshift=-0.2cm] plot [smooth] coordinates  {(0,0) (4,0)};
    \draw[-,      very thick, CUDgray,   yshift= 0.2cm] plot [smooth] coordinates  {(0,0) (4,0)};

    \draw[-latex, very thick, CUDred, yshift=0.08cm] plot [smooth] coordinates {(0,0) (4,0)};

    \node[anchor=east] at (10,0)
      {$X$-string (red:\,$W^3$-flux, navy:\,$A$-flux)};
  \end{scope}
\end{tikzpicture}
    \caption{Schematic illustrations of the magnetic fluxes inside the $Z$- and $X$-strings, as described in Eqs.\,\eqref{eq:Z-potential as WB-potential} and \eqref{eq:X-potential as WB-potential}. The magnetic $Z$-flux and $X$-flux are expressed in terms of the corresponding magnetic $W^3$- and $A$-fluxes.
    }
    \label{fig:ZandXstrings}
\end{figure}
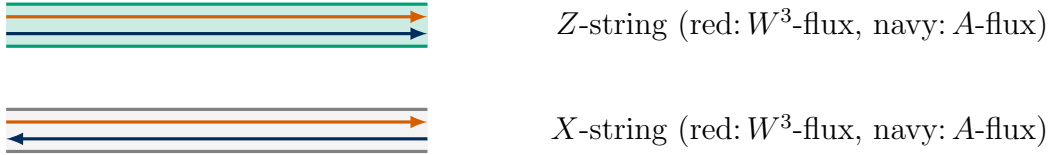

\subsubsection*{\texorpdfstring{$\boldsymbol{Z}$}{Z}-string}

The asymptotic behavior of strings formed at $\mathrm{U}(1)_\mathrm{M}\times \mathrm{U}(1)_\mathrm{P} \to \mathrm{U}(1)_\mathrm{X}$ is given by
\begin{align}
    H \to \VH\, e^{i n \varphi} \ , \quad
    Z \to \frac{2n}{g_Z} \dd \varphi\ ,
    \label{eq:Z-potential}
\end{align}
for $\rho \gg m_Z^{-1}$, while $X_\mu = 0$. Here, we define $g_Z = \sqrt{g^2 + g'^2}$. 
The magnetic $Z$-flux runs along the $Z$-string.
For later purposes, let us express the asymptotic behavior of the $Z$-string in terms of the original gauge fields. Using the mixing matrix in Eq.\,\eqref{eq:mixing}, we find
\begin{align}
    W^3 \to \frac{2n}{g} c_\alpha^2 \dd \varphi\ , \quad 
    A \to \frac{2n}{g'} s_\alpha^2 \dd \varphi\ .
    \label{eq:Z-potential as WB-potential}
\end{align}
The magnetic fluxes in terms of the original gauge fields $W^3$ and $A$ along the $Z$-string are
\begin{alignat}{2}
\int \dd[2]{x}\, g W^3_{12} &= \oint_{\rho\rightarrow \infty} g W^3_i\, \dd{x}^i & &= 4\pi n\, c_\alpha^2\ , \\
\int \dd[2]{x}\, g' F_{12}   &= \oint_{\rho\rightarrow \infty} g' A_i\, \dd{x}^i & &= 4\pi n\, s_\alpha^2\ .
\label{eq:Z-flux as WB-flux}
\end{alignat}
Here, $F_{\mu\nu}$ is the field strength of $A_\mu$.

\subsubsection*{\texorpdfstring{$\boldsymbol{X}$}{X}-string}
The asymptotic behavior of strings formed at $\mathrm{U}(1)_\mathrm{X}\to \mathrm{Nothing}$ is given by
\begin{align}
    \eta \to v_\eta\, e^{i n_\eta \varphi}\ , \quad
    X \to \frac{1}{g_X}\frac{n_\eta}{ Q_\eta} \dd \varphi\ ,
    \label{eq:X-potential}
\end{align}
with $n_\eta \in \mathbbm{Z}$ 
for $\rho \gg m_X^{-1}$, while $Z = 0$. This isolated $X$-string corresponds to a stable ANO-type cosmic string.
In terms of the original gauge fields, $W^3$ and $A$, the asymptotic behavior of $X$-string corresponds to
\begin{align}
    \label{eq:X-potential as WB-potential}
    W^3 \to - \frac{s_\alpha}{g_X} \frac{n_\eta}{Q_\eta} \dd \varphi
    = - \frac{1}{g} \frac{n_\eta}{Q_\eta} \dd \varphi\ , \quad 
    A \to \frac{c_\alpha}{g_X} \frac{n_\eta}{Q_\eta} \dd \varphi
    = \frac{1}{g'} \frac{n_\eta}{Q_\eta} \dd \varphi\ .
\end{align}
The magnetic fluxes along the $X$-string in terms of the original gauge fields are given by
\begin{alignat}{2}
\label{eq: string flux 3}
\int \dd[2]{x}\, g W^3_{12} &= \oint_{\rho\rightarrow \infty} g W^3_i\, \dd{x}^i & &= -\frac{2\pi n_\eta}{Q_\eta}\ , \\
\int \dd[2]{x}\, g' F_{12}   &= \oint_{\rho\rightarrow \infty} g' A_i\, \dd{x}^i & &= \frac{2\pi n_\eta}{Q_\eta}\ .
\end{alignat}
In Fig.\,\ref{fig:ZandXstrings}, we show 
schematic pictures of the $Z$-string and $X$-string.
The magnetic $Z$-flux ($X$-flux) flows at the core 
of the $Z$-string ($X$-string).
In the figure, we present them as the magnetic flux of the original gauge fields 
$W^3$ and $A$.

\subsection{Case A : String Decay Before \texorpdfstring{$\boldsymbol{\mathrm{U(1)}_\mathrm{X}}$}{U(1)X} Symmetry Breaking}
In Case A, we consider the situation where string decay occurs before $\mathrm{U}(1)_\mathrm{X}\to \mathrm{Nothing}
$ breaking. 
As shown in Eq.\,\eqref{eq: monopole flux}, the magnetic $W$-flux from a monopole is $-4\pi$. 
On the other hand, the magnetic $W^3$-flux carried by the $Z$-string is $4\pi c_\alpha^2$ (see Eq.\,\eqref{eq:Z-flux as WB-flux}). 
Therefore, unlike in the simple SU(2) model discussed in Sec.\,\ref{sec:monopole connection}, 
there is a mismatch between the monopole's flux and the string flux.
Accordingly, when the monopole is attached to the $Z$-string, a part of the monopole's magnetic $W^3$-flux is not confined within the string but instead leaks into the surrounding space,
\begin{align}
\int_{(S^2\setminus\mathrm{str})_\infty} g\, \dd W^3
= -4\pi + 4\pi c_\alpha^2 = -4\pi s_\alpha^2\ ,
\label{eq:emitted in fluxW}
\end{align}
where $(S^2\setminus\mathrm{str})_\infty$ denotes an infinitely large  spherical surface
surrounding the monopole with the 
interior of the
$Z$-string excluded.
The radius of the string is taken to be finite but much larger than $m_{X,Z}^{-1}$.
Here, the orientation of the $Z$-string 
is arranged such that
the magnetic $W^3$-flux in the string is 
\begin{align}
\label{eq:Z-string flux}
\int_{\mathrm{str}} g\, \dd W^3 = -4\pi c_\alpha^2 \ .
\end{align}
In addition to the $W^3$-flux, the $Z$-string also carries magnetic $A$-flux, as described in Eq.\,\eqref{eq:Z-potential as WB-potential}. 
Since the $A$-flux is associated with $\mathrm{U}(1)_\mathrm{P}$ , it cannot terminate anywhere, including at the core of the monopole. 
Consequently, the $A$-flux inside the $Z$-string must flow out through the monopole,
\begin{align}
\int_{(S^2\setminus \mathrm{str})_\infty} g'\,\dd A
= -\int_{\mathrm{str}} g'\,\dd A
= 4\pi s_\alpha^2\ ,
\label{eq:flowing out fluxB}
\end{align}
where the negative sign arises from the orientation of the $Z$-string.
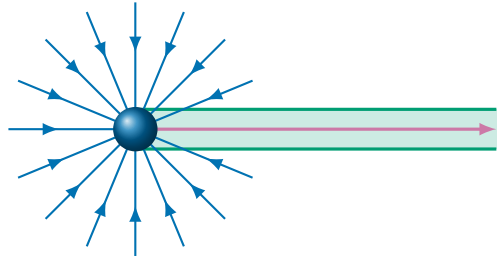
\begin{figure}[t]
    \centering
\begin{tikzpicture}[
    scale=2,
    >=latex,
    arrow inside/.style={
      postaction={decorate},
      decoration={
        markings,
        mark=at position 0.4 with {\arrow[scale=1.2]{>}}
      }
    }
]
  \begin{scope}[scale=1.2]
    \fill[CUDgreen, opacity=0.2] (0,-0.11) rectangle (2,0.11);
    \draw[-latex, very thick, CUDpurple] plot [smooth] coordinates {(0,0) (2,0)};
    \draw[-,      very thick, CUDgreen, yshift=-0.11cm] plot [smooth] coordinates {(0,0) (2,0)};
    \draw[-,      very thick, CUDgreen, yshift= 0.11cm] plot [smooth] coordinates {(0,0) (2,0)};

    \foreach \i in {22.5,45,...,337.5} {
      \draw[arrow inside, thick, CUDblue] (\i:20pt) -- (0,0);
    }

    \shade[ball color=CUDblue] (0,0) circle (3.5pt);
  \end{scope}
\end{tikzpicture}
    \caption{A schematic picture of a monopole attached to a cosmic string in Case A, where string decay occurs before the $\mathrm{U}(1)_\mathrm{X}$ symmetry breaking. The purple arrow represents the confined magnetic $Z$-flux, while the blue arrows indicate the unconfined magnetic $X$-flux.
    An antimonopole similarly emits magnetic fluxes with opposite orientation.}
    \label{fig:attached2}
\end{figure}
The existence of magnetic flux outside the $Z$-string segment in Eqs.\,\eqref{eq:emitted in fluxW} and
\eqref{eq:flowing out fluxB} indicates the presence of unconfined $X$-flux,
\begin{align}
\int_{(S^2\setminus \mathrm{str})_\infty} \dd X = 
\frac{4\pi s_\alpha^2}{g_X}\ ,
\label{eq:X-flux around monopole}
\end{align}
while
\begin{align}
\int_{(S^2\setminus \mathrm{str})_\infty} \dd Z = 0\ ,
\end{align}
in the region outside the monopole (see Eqs.\,\eqref{eq:mixing}). Since the $\mathrm{U}(1)_\mathrm{X}$ symmetry remains unbroken, this $X$-flux is not confined and instead spreads out spherically. 
As a result, in Case A, the monopole is accompanied by a spherically spreading magnetic $X$-flux, while the $Z$-flux is confined within the cosmic string. A schematic illustration of this configuration is shown in Fig.\,\ref{fig:attached2}.

\begin{figure}
    \centering
 \begin{tikzpicture}[scale=2]
  \begin{scope}
    \fill[CUDgreen, opacity=0.2] (0,-0.11) rectangle (4,0.11);
    \draw[-latex, very thick, CUDpurple] plot [smooth] coordinates {(0,0) (4,0)};
    \draw[-,      very thick, CUDgreen, yshift=-0.11cm] plot [smooth] coordinates {(0,0) (4,0)};
    \draw[-,      very thick, CUDgreen, yshift= 0.11cm] plot [smooth] coordinates {(0,0) (4,0)};
  \end{scope}

  \draw[line width=0.4mm, -{Stealth[length=4mm, round]}] (2,-0.3) -- (2,-0.8);
\end{tikzpicture}

\vspace{.5em}

\begin{tikzpicture}[scale=2]
  \begin{scope}
    \fill[CUDgreen, opacity=0.2] (0,-0.11) rectangle (2,0.11);
    \draw[-latex, very thick, CUDpurple] plot [smooth] coordinates {(0,0) (1.9,0)};
    \draw[-,      very thick, CUDgreen,   yshift=-0.11cm] plot [smooth] coordinates {(0,0) (2,0)};
    \draw[-,      very thick, CUDgreen,   yshift= 0.11cm] plot [smooth] coordinates {(0,0) (2,0)};
    \shade[ball color=CUDorange] (2,0) circle (3.5pt);
  \end{scope}

  \begin{scope}[xshift=3cm]
    \fill[CUDgreen, opacity=0.2] (0,-0.11) rectangle (2,0.11);
    \draw[-latex, very thick, CUDpurple] plot [smooth] coordinates {(0,0) (2,0)};
    \draw[-,      very thick, CUDgreen,   yshift=-0.11cm] plot [smooth] coordinates {(0,0) (2,0)};
    \draw[-,      very thick, CUDgreen,   yshift= 0.11cm] plot [smooth] coordinates {(0,0) (2,0)};
    \shade[ball color=CUDblue] (0,0) circle (3.5pt);
  \end{scope}

  \draw[-latex, very thick, CUDblue]
    (2,  0.15) to[out=80,  in=110] (3,  0.15);
  \draw[-latex, very thick, CUDblue]
    (2.1,0.15) to[out=30,  in=150] (2.9,0.15);
  \draw[-latex, very thick, CUDblue]
    (2.15,0.00) to[out=0,   in=180] (2.85,0.00);
  \draw[-latex, very thick, CUDblue]
    (2.1,-0.15) to[out=-30, in=210] (2.9,-0.15);
  \draw[-latex, very thick, CUDblue]
    (2,-0.15) to[out=-80, in=250] (3,-0.15);
\end{tikzpicture}
    \caption{Decay of the cosmic string by monopole–antimonopole pair nucleation in Case A. The magnetic $Z$-flux (purple) is confined within the string, while the magnetic $X$-flux (blue) is unconfined and spreads between the pair.}
    \label{fig:string breaking2}
\end{figure}
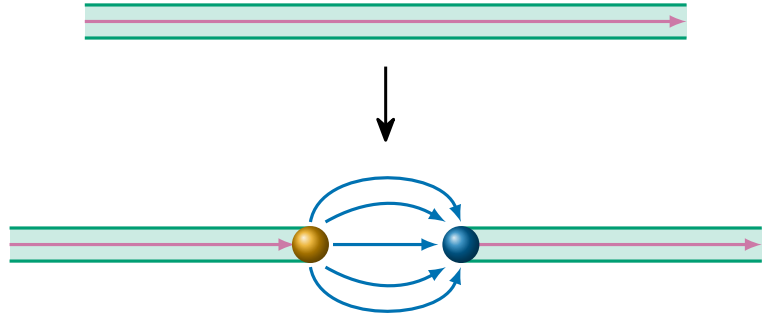

Let us now consider the decay process of the cosmic string. In Case A, where the $\mathrm{U}(1)_\mathrm{X}$ symmetry remains unbroken, a magnetic $X$-flux surrounds the monopole, as discussed above. When a monopole–antimonopole pair is nucleated within a long string, the $X$-flux lines connect the pair (see Fig.\,\ref{fig:string breaking2}). 
The presence of the $X$-flux between the monopole-antimonopole pair significantly affects the subsequent evolution of the resulting string segments. In particular, segments carrying unconfined $X$-flux can radiate
the massless $\U(1)_\mathrm{X}$ gauge boson and can
interact with $\mathrm{U}(1)_\mathrm{X}$-charged particles in the thermal plasma.
Through these interactions, the energy stored in the string segments is transferred to the plasma, although we do not explore this process in detail here.

\subsection{Case B : String Decay After \texorpdfstring{$\boldsymbol{\mathrm{U(1)}_\mathrm{X}}$}{U(1)X} Symmetry Breaking}
Let us discuss the behavior of $\eta$ in the presence of a $Z$-string which occurs in Case B.
The kinetic terms of $H_1$ and $\eta$ are given by
\begin{align}
\mathcal{L}_\mathrm{kin} =
\left| \left(\partial_\mu - \frac{i}{2} g W^3_\mu - \frac{i}{2} g' A_\mu \right) H_1 \right|^2
+
\left| \left(\partial_\mu - i Q_\eta g' A_\mu \right) \eta \right|^2\ .
\end{align}
In the presence of a $Z$-string, $H_1$ behaves asymptotically as $H_1 \to v_H e^{i\varphi}$ for $\rho\to \infty$. 
We assume that $\eta$ has a winding number $n_\eta$, such that $\eta \to v_\eta e^{i n_\eta \varphi}$
for $\rho\to \infty$.
With these assumptions, the asymptotic behavior of the gauge potentials for $\rho \to \infty$ 
is given by
\begin{align}
W^3 \to \frac{1}{g} \left(2 - \frac{n_\eta}{Q_\eta}\right) \mathrm{d}\varphi, \quad
A \to \frac{1}{g'} \frac{n_\eta}{Q_\eta} \mathrm{d}\varphi\ .
\label{eq: WA boundary condition}
\end{align}
These correspond to the following asymptotic forms for $Z$ and $X$,
\begin{align}
Z \to \frac{2}{g_Z} \mathrm{d}\varphi\ , \quad
X \to \frac{1}{g_X} \left(\frac{n_\eta}{Q_\eta} - 2 s_\alpha^2 \right) \mathrm{d}\varphi\ .
\label{eq:asymptotic Z and X}
\end{align}
We refer to a configuration with these asymptotic forms for $H_1$ and $\eta$ as a $ZX$-string.

\begin{figure}
    \centering
\begin{tikzpicture}[scale=1.4]
  \begin{scope}
    \fill[CUDgray, opacity=0.09] (0,0.2) rectangle (4,0.6);
    \fill[CUDgray, opacity=0.09] (0,-0.2) rectangle (4,-0.6);
    \fill[CUDgreen, opacity=0.2] (0,-0.2) rectangle (4,0.2);

    \draw[-,        very thick, CUDgray,  yshift=-0.6cm] plot [smooth] coordinates {(0,0) (4,0)};
    \draw[-,        very thick, CUDgray,  yshift= 0.6cm] plot [smooth] coordinates {(0,0) (4,0)};
    \draw[-,        very thick, CUDgreen, yshift=-0.2cm] plot [smooth] coordinates {(0,0) (4,0)};
    \draw[-,        very thick, CUDgreen, yshift= 0.2cm] plot [smooth] coordinates {(0,0) (4,0)};

    \draw[-latex,   very thick, CUDpurple,yshift=-0.08cm] plot [smooth] coordinates {(0,0.1) (4,0.1)};
    \draw[latex-,   very thick, CUDblue,  yshift=-0.5cm] plot [smooth] coordinates {(0,0.1) (4,0.1)};
    \draw[latex-,   very thick, CUDblue,  yshift= 0.5cm] plot [smooth] coordinates {(0,-0.1) (4,-0.1)};

    \draw[<->, very thick, black] plot [smooth] coordinates {(4.2,-0.2) (4.2,0.2)};
    \node at (4.7,0) {$m_Z^{-1}$};

    \draw[<->, very thick, black] plot [smooth] coordinates {(-0.2,-0.6) (-0.2,0.6)};
    \node at (-0.6,0) {$m_X^{-1}$};
  \end{scope}
\end{tikzpicture}
    \caption{Schematic illustration of the magnetic flux inside a $ZX$-string with $n_\eta = 0$ in Case B. 
The purple and blue arrows represent the $Z$- and $X$-fluxes, respectively, as given in Eq.\,\eqref{eq: flux in ZX-string}. 
The $Z$-flux is confined within $\rho \lesssim m_Z^{-1}$, while the $X$-flux is confined within $\rho \lesssim m_X^{-1}$.
    }
    \label{fig:ZXstrings}
\end{figure}
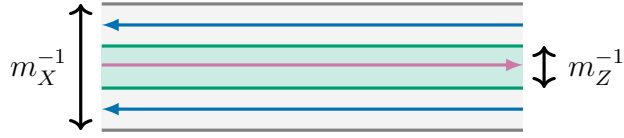

As we will see in Sec.\,\ref{sec:numerical},
the magnetic $Z$ field is localized within $\rho \lesssim m_Z^{-1}$, while the magnetic $X$ field resides in the region $\rho \lesssim m_X^{-1}$. The total magnetic $Z$- and $X$-fluxes are fixed by the asymptotic behavior in Eq.\,\eqref{eq:ZX asymptotic}. 
Thus, for $n_\eta =0 $, 
the $ZX$-string carries magnetic fluxes
\begin{align}
\label{eq: flux in ZX-string}
    \int_\mathrm{str} \dd Z = \frac{4\pi}{g_Z}\ ,\qquad
    \int_\mathrm{str} \dd X = 
    -\frac{4\pi s_\alpha^2}{g_X}\ ,
\end{align}
confined within $\rho \lesssim m_Z^{-1}$ and $\rho \lesssim m_X^{-1}$, respectively.
In terms of the original gauge fields $W^3$ and $A$, these fluxes correspond to
\begin{align}
\label{eq: W3 and B in ZX-string}
    \int_\mathrm{str} \dd W^3 = \frac{4\pi}{g}\ , \qquad
    \int_\mathrm{str} \dd A = 0\ ,
\end{align}
for $n_\eta = 0$.
Thus, 
when $n_\eta = 0$, the total magnetic $W^3$-flux confined in the $ZX$-string (as given in Eq.\,\eqref{eq: W3 and B in ZX-string}) precisely matches the magnetic $W$-flux carried by the monopole.
For $n_\eta \neq 0$, the $ZX$-string 
carries additional magnetic $X$-flux
which can be read from the asymptotic behavior in Eq.\,\eqref{eq:asymptotic Z and X}.
A schematic illustration of the $ZX$-string configuration is shown in Fig.\,\ref{fig:ZXstrings}.

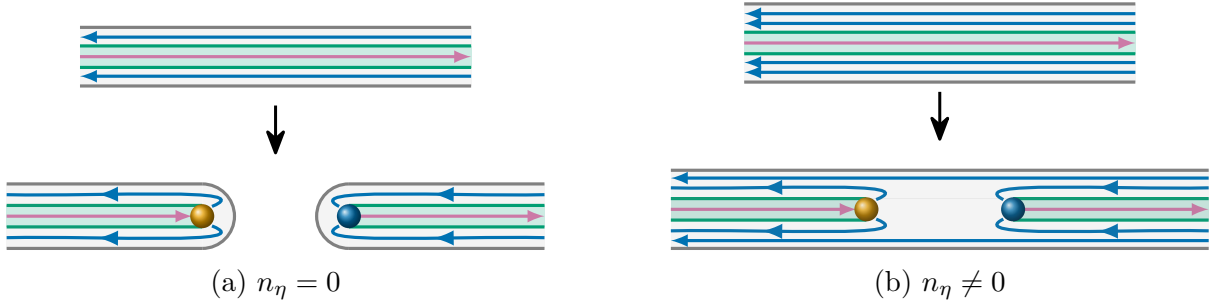
\begin{figure}[t]
\centering
\begin{subfigure}{0.45\textwidth}
    \centering
\begin{tikzpicture}[scale=1.3,>=latex]
  \begin{scope}[yshift=0.2cm]
    \fill[CUDgray, opacity=0.09] (0,0.11) rectangle (4,0.3);
    \fill[CUDgray, opacity=0.09] (0,-0.11) rectangle (4,-0.3);
    \draw[latex-, very thick, CUDblue] plot [smooth] coordinates {(0,0.2) (4,0.2)};
    \draw[latex-, very thick, CUDblue] plot [smooth] coordinates {(0,-0.2) (4,-0.2)};
    \fill[CUDgreen, opacity=0.2] (0,-0.11) rectangle (4,0.11);
    \draw[-latex, very thick, CUDpurple] plot [smooth] coordinates {(0,0) (4,0)};
    \draw[-, very thick, CUDgray,yshift=-0.3cm] plot [smooth] coordinates {(0,0) (4,0)};
    \draw[-, very thick, CUDgray,yshift=0.3cm] plot [smooth] coordinates {(0,0) (4,0)};
    \draw[-, very thick, CUDgreen,yshift=-0.11cm] plot [smooth] coordinates {(0,0) (4,0)};
    \draw[-, very thick, CUDgreen,yshift=0.11cm] plot [smooth] coordinates {(0,0) (4,0)};
  \end{scope}
  \draw[line width=0.4mm, -{Stealth[length=4mm, round]}] (2,-0.3) -- (2,-0.8);
\end{tikzpicture}

\vspace{.3cm}

\begin{tikzpicture}[scale=1.3]
  \begin{scope}
    \fill[CUDgreen, opacity=0.2] (0,-0.11) rectangle (2,0.11);
    \draw[-latex, very thick, CUDpurple] plot [smooth] coordinates {(0,0) (1.9,0)};
    \draw[-, very thick, CUDgreen,yshift=-0.11cm] plot [smooth] coordinates {(0,0) (2,0)};
    \draw[-, very thick, CUDgreen,yshift=0.11cm] plot [smooth] coordinates {(0,0) (2,0)};
    \draw[CUDgray, very thick] (2.0,0.33) arc[start angle=90, end angle=-90, radius=.33]; 
    \fill[CUDgray, opacity=0.09] (2.0,0.33) arc[start angle=90, end angle=-90, radius=.33] -- cycle;    
    \fill[CUDgray, opacity=0.09] (0.,0.10) rectangle (2,0.33);
    \fill[CUDgray, opacity=0.09] (0,-0.10) rectangle (2,-0.33);
    \shade[ball color=CUDorange] (2,0) circle (3.5pt);
    \draw[-, very thick, CUDgray,yshift=0.33cm] plot [smooth] coordinates {(0,0) (2,0)};
    \draw[-, very thick, CUDgray,yshift=-0.33cm] plot [smooth] coordinates {(0,0) (2,0)};

    \draw[
      very thick,
      CUDblue,
      postaction={decorate},
      decoration={
        markings,
        mark=at position 0.6 with {\arrow[scale=1.2]{latex}}
      }
    ] (2.1,-0.1) .. controls (2.6,-.3) and (1.0,-0.2) .. (0,-0.22);
    \draw[
      very thick,
      CUDblue,
      postaction={decorate},
      decoration={
        markings,
        mark=at position 0.6 with {\arrow[scale=1.2]{latex}}
      }
    ] (2.1,0.1) .. controls (2.6,.3) and (1.0,0.2) .. (0,0.22);
  \end{scope}

  \begin{scope}[xshift=3.5cm]
    \draw[CUDgray, very thick] (0.0,0.33) arc[start angle=90, end angle=270, radius=.33];
    \fill[CUDgray, opacity=0.09] (0.0,0.33) arc[start angle=90, end angle=270, radius=.33] -- cycle;
    \fill[CUDgray, opacity=0.09] (0.0,0.10) rectangle (2,0.33);
    \fill[CUDgray, opacity=0.09] (0.0,-0.10) rectangle (2,-0.33);
    \fill[CUDgreen, opacity=0.2] (0,-0.11) rectangle (2,0.11);
    \draw[-latex, very thick, CUDpurple] plot [smooth] coordinates {(0,0) (2,0)};
    \draw[-, very thick, CUDgray,yshift=0.33cm] plot [smooth] coordinates {(0,0) (2,0)};
    \draw[-, very thick, CUDgray,yshift=-0.33cm] plot [smooth] coordinates {(0,0) (2,0)};
    \draw[-, very thick, CUDgreen,yshift=-0.11cm] plot [smooth] coordinates {(0,0) (2,0)};
    \draw[-, very thick, CUDgreen,yshift=0.11cm] plot [smooth] coordinates {(0,0) (2,0)};
    \shade[ball color=CUDblue] (0,0) circle (3.5pt);
    \draw[
      very thick,
      CUDblue,
      postaction={decorate},
      decoration={
        markings,
        mark=at position 0.5 with {\arrow[scale=1.2]{latex}}
      }
    ] (2,-0.22) .. controls (1.0,-0.2) and (-.5,-.3) .. (-0.1,-0.1);
    \draw[
      very thick,
      CUDblue,
      postaction={decorate},
      decoration={
        markings,
        mark=at position 0.5 with {\arrow[scale=1.2]{latex}}
      }
    ] (2,0.22) .. controls (1.0,0.2) and (-.5,.3) .. (-0.1,0.1);
  \end{scope}
\end{tikzpicture}    
\caption{$n_\eta = 0$}
 \label{fig:string breaking3}
 \end{subfigure}
\hspace{.5cm}
\begin{subfigure}{0.45\textwidth}
    \centering
\begin{tikzpicture}[scale=1.3,>=latex]
  \begin{scope}[yshift=0.2cm]
    \fill[CUDgray, opacity=0.09] (0,0.11) rectangle (4,0.4);
    \fill[CUDgray, opacity=0.09] (0,-0.11) rectangle (4,-0.4);
    \draw[latex-, very thick, CUDblue] plot [smooth] coordinates {(0,0.3) (4,0.3)};
    \draw[latex-, very thick, CUDblue] plot [smooth] coordinates {(0,0.2) (4,0.2)};
    \draw[latex-, very thick, CUDblue] plot [smooth] coordinates {(0,-0.2) (4,-0.2)};
      \draw[latex-, very thick, CUDblue] plot [smooth] coordinates {(0,-0.3) (4,-0.3)};
    \fill[CUDgreen, opacity=0.2] (0,-0.11) rectangle (4,0.11);
    \draw[-latex, very thick, CUDpurple] plot [smooth] coordinates {(0,0) (4,0)};
    \draw[-, very thick, CUDgray,yshift=-0.4cm] plot [smooth] coordinates {(0,0) (4,0)};
    \draw[-, very thick, CUDgray,yshift=0.4cm] plot [smooth] coordinates {(0,0) (4,0)};
    \draw[-, very thick, CUDgreen,yshift=-0.11cm] plot [smooth] coordinates {(0,0) (4,0)};
    \draw[-, very thick, CUDgreen,yshift=0.11cm] plot [smooth] coordinates {(0,0) (4,0)};
  \end{scope}

  \draw[line width=0.4mm, -{Stealth[length=4mm, round]}] (2,-0.3) -- (2,-0.8);
\end{tikzpicture}

\vspace{.3cm}

\begin{tikzpicture}[scale=1.3]
  \begin{scope}
    \fill[CUDgreen, opacity=0.2] (0,-0.11) rectangle (2,0.11);
    \draw[-latex, very thick, CUDpurple] plot [smooth] coordinates {(0,0) (1.9,0)};
    \draw[-, very thick, CUDgreen,yshift=-0.11cm] plot [smooth] coordinates {(0,0) (2,0)};
    \draw[-, very thick, CUDgreen,yshift=0.11cm] plot [smooth] coordinates {(0,0) (2,0)};
    \shade[ball color=CUDorange] (2,0) circle (3.5pt);
    \draw[
      very thick,
      CUDblue,
      postaction={decorate},
      decoration={
        markings,
        mark=at position 0.6 with {\arrow[scale=1.2]{latex}}
      }
    ] (2.1,-0.1) .. controls (2.6,-.3) and (1.0,-0.2) .. (0,-0.22);
    \draw[
      very thick,
      CUDblue,
      postaction={decorate},
      decoration={
        markings,
        mark=at position 0.6 with {\arrow[scale=1.2]{latex}}
      }
    ] (2.1,0.1) .. controls (2.6,.3) and (1.0,0.2) .. (0,0.22);
  \end{scope}

  \begin{scope}[xshift=3.5cm]
    \fill[CUDgreen, opacity=0.2] (0,-0.11) rectangle (2,0.11);
    \draw[-latex, very thick, CUDpurple] plot [smooth] coordinates {(0,0) (2,0)};
    \draw[-, very thick, CUDgreen,yshift=-0.11cm] plot [smooth] coordinates {(0,0) (2,0)};
    \draw[-, very thick, CUDgreen,yshift=0.11cm] plot [smooth] coordinates {(0,0) (2,0)};
    \shade[ball color=CUDblue] (0,0) circle (3.5pt);
    \draw[
      very thick,
      CUDblue,
      postaction={decorate},
      decoration={
        markings,
        mark=at position 0.5 with {\arrow[scale=1.2]{latex}}
      }
    ] (2,-0.22) .. controls (1.0,-0.2) and (-.5,-.3) .. (-0.1,-0.1);
    \draw[
      very thick,
      CUDblue,
      postaction={decorate},
      decoration={
        markings,
        mark=at position 0.5 with {\arrow[scale=1.2]{latex}}
      }
    ] (2,0.22) .. controls (1.0,0.2) and (-.5,.3) .. (-0.1,0.1);
  \end{scope}

  \begin{scope}
    \fill[CUDgray, opacity=0.09] (0,0.11) rectangle (5.5,0.4);
    \fill[CUDgray, opacity=0.09] (0,-0.11) rectangle (5.5,-0.4);
    \fill[CUDgray, opacity=0.09] (0,0.11) rectangle (5.5,-0.11);
    \draw[-, very thick, CUDgray] plot [smooth] coordinates {(0,0.4) (5.5,0.4)};
    \draw[-, very thick, CUDgray] plot [smooth] coordinates {(0,-0.4) (5.5,-0.4)};
    \draw[latex-, very thick, CUDblue] plot [smooth] coordinates {(0,-0.32) (5.5,-0.32)};
    \draw[latex-, very thick, CUDblue] plot [smooth] coordinates {(0,0.32) (5.5,0.32)};
  \end{scope}
\end{tikzpicture}
\caption{$n_\eta \neq 0$}
\label{fig:string breaking4}
\end{subfigure}
    \caption{Decay of the cosmic string by monopole–antimonopole pair nucleation in Case B with 
   (a) $n_\eta = 0$ 
   and (b)
    $n_\eta \neq 0$. 
The magnetic $Z$-flux (purple) is confined within $\rho \lesssim m_Z^{-1}$, while the magnetic $X$-flux (blue) is confined within $\rho \lesssim m_X^{-1}$. 
After the string decay, no $X$-string remains stretched between the monopole and antimonopole for $n_\eta = 0$. For $n_\eta \neq 0$, 
on the other hand, 
only $X$-string is left after
the monopole-antimonopole pair nucleation. 
}
\end{figure}

Finally, let us discuss the string decay process of the cosmic string in Case B. 
For $n_\eta = 0$, 
the total magnetic $W$- and $A$-fluxes confined in the $ZX$-string (as given in Eq.\,\eqref{eq: W3 and B in ZX-string}) matches the monopole magnetic flux.
Therefore, 
when a monopole–antimonopole pair is nucleated within the string, the magnetic $Z$- and $X$-fluxes in the cosmic string are smoothly connected to the monopole and antimonopole without any mismatch. 
In Fig.\,\ref{fig:string breaking3}, we show a schematic picture of the string decay in Case B with $n_\eta = 0$.
In this case, there is neither magnetic flux nor winding of $\eta$ between the monopole and antimonopole, resulting in a topologically trivial configuration. As a result, the broken string segment in this case resembles that in the simple SU(2) model and carries only confined flux.

For $n_\eta \neq 0$, on the other hand, a portion of the $X$-flux cannot be absorbed by the monopoles. Consequently, after the string decay, an $X$-string is left after $ZX$-string decay (see Fig.\,\ref{fig:string breaking4}).
In the regime where $v_H \gg v_\eta$, the cosmic strings that predominantly contribute to the gravitational wave signal are the $Z$-strings and $ZX$-strings, whose string tensions are of order $\mathcal{O}(v_H^2)$. In contrast, the contribution from $X$-strings, which have much lower tension of $\order{v_\eta^2}$, is negligible.

\subsection{Radial Configuration of \texorpdfstring{$\boldsymbol{ZX}$}{ZX}-String}
\label{sec:numerical}
In the previous subsection, we 
discussed the decay of the $ZX$-string by assuming that the magnetic $Z$ field is localized within $\rho \lesssim m_Z^{-1}$, while the magnetic $X$ field resides in the region $\rho \lesssim m_X^{-1}$.
In this subsection,
we confirm this behavior by 
analyzing the radial configuration
of $ZX$-string numerically.

For this purpose, we introduce the following profile functions,
\begin{align}
H_1 = v_H e^{i \varphi} F_H(\rho)\ , \quad
\eta = v_\eta e^{i n_\eta \varphi} F_\eta(\rho)\ , \quad
W^3 = \frac{1}{g} f_W(\rho) \mathrm{d}\varphi\ , \quad
A = \frac{1}{g'} f_A(\rho) \mathrm{d}\varphi\ .
\label{eq:ansatz}
\end{align}
The corresponding profile functions 
of $Z$ and $X$ are given by,
\begin{align}
f_Z(\rho) &= g_Z\left(c_\alpha \frac{f_W(\rho)}{g} + s_\alpha \frac{f_A(\rho)}{g'}\right)\ , \\
f_X(\rho) &= g_X\left(-s_\alpha \frac{f_W(\rho)}{g} + c_\alpha \frac{f_A(\rho)}{g'}\right)\ ,
\end{align}
(see Eq.\,\eqref{eq:mixing})
The corresponding gauge fields are,
\begin{align}
\label{eq:ZX asymptotic}
Z = \frac{1}{g_Z} f_Z(\rho) \mathrm{d}\varphi, \quad X = \frac{1}{g_X} f_X(\rho) \mathrm{d}\varphi\ ,
\end{align}
respectively.
All the profile functions are taken to be real.

The equations of motion for these profile functions are,
\begin{align}
&\frac{(\rho F_H')'}{\rho}
- \frac{1}{\rho^2}\qty(1 - \frac{f_Z}{2})^2 F_H - 2\lambda_H v_H^2 (F_H^2 - 1) F_H = 0\
 ,\\
&\frac{(\rho F_\eta')'}{\rho}- \frac{1}{\rho^2}(n_\eta - Q_\eta (f_X+s_\alpha^2f_Z))^2 F_\eta - 2\lambda_\eta v_\eta^2 (F_\eta^2 - 1) F_\eta = 0\ , \label{eq:FetaEoM}\\
&\rho\left(\frac{f_Z'}{\rho}\right)' + {g_Z^2 v_H^2 F_H^2}\left(1 - \frac{f_Z}{2} \right) 
+2s_\alpha^2 g_Z^2  Q_\eta^2 v_\eta^2 F_\eta^2(n_\eta-Q_\eta (f_X +s_\alpha^2 f_Z)) 
= 0\ ,\\
&\rho\left(\frac{f_X'}{\rho}\right)' 
- {2g_X^2 Q_\eta^2 v_\eta^2 F_\eta^2}(n_\eta-Q_\eta (f_X +s_\alpha^2 f_Z)) = 0\ .
\end{align}
Here, the prime denotes differentiation with respect to $\rho$. The boundary conditions are given by:
\begin{gather}
F_H(0) = 0\ , \quad  F_H(\infty) = 1\ , \quad
F_\eta(0) = 0\ , \quad F_\eta(\infty) = 1\ , \\
f_Z(0) = 0\ , \quad f_Z(\infty) = 2 \ , \quad
f_X(0) = 0\ , \quad f_X(\infty) = \frac{n_\eta}{Q_\eta} - 2 s_\alpha^2 \ , \label{eq:ZX boundary condition}
\end{gather}
for $n_\eta \neq 0$.
For $n_\eta = 0$, $F'_\eta(0)=0$ follows from Eq.\,\eqref{eq:FetaEoM}.

\begin{figure}[t]
\centering  
\fcolorbox{blue}{white}{\large{${v_\eta/v_H=0.1}$}}
\includegraphics[width=0.85\textwidth]{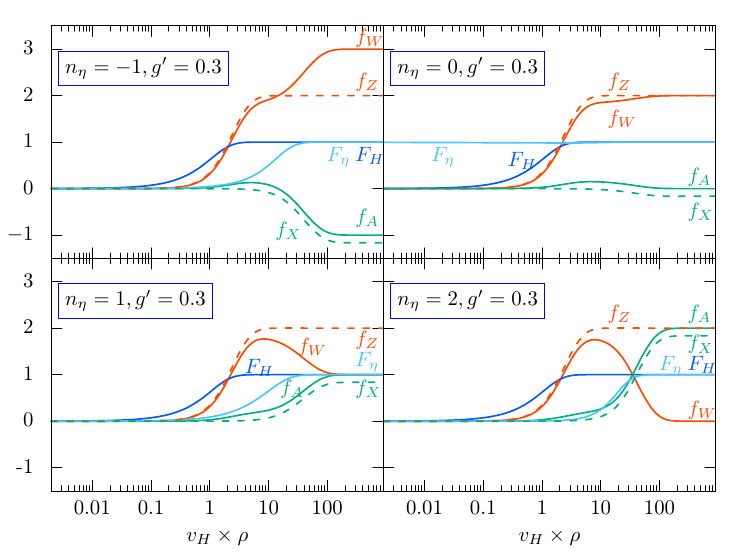}
\caption{Profile functions 
of the $ZX$-string for the given parameters. 
The scalar fields are wound like $H_1 \propto e^{i\varphi}$ and $\eta \propto e^{i n_\eta \varphi}$, with $n_\eta = -1$, $0$, $1$ and $2$. 
The blue, cyan, red, and green solid lines represent $F_H$, $F_\eta$, $f_W$, and $f_A$, respectively. The red and green dashed lines indicate $f_Z$ and $f_X$, respectively.}
  \label{fig:profiles}
\end{figure}

The profile functions for the $ZX$-string are shown in Fig.\,\ref{fig:profiles} for the parameter set
\begin{align}
g = 1\ , \quad g' = 0.3\ , \quad \lambda_H = \lambda_\eta = \frac{1}{2}\ , \quad v_\eta = 0.1\times v_H\ ,  \quad Q_\eta = 1\ .
\label{eq:sample}
\end{align}
This corresponds to the mass spectrum
\begin{align}
\frac{m_H}{v_H} \simeq 1.4\ , \quad \frac{m_Z}{v_H} \simeq 0.74\ , \quad \frac{m_\eta}{v_H} \simeq 0.14\ , \quad \frac{m_X}{v_H} \simeq 0.04\ .
\end{align}
The winding number is set to $n_\eta = -1$, $0$, $1$ and $2$. 

The figure shows that $F_H(\rho)$ transitions from $0$ to $1$ over $\rho \sim m_H^{-1}$, and similarly $F_\eta(\rho)$ for $n_\eta \neq 0$ transitions over $\rho \sim m_\eta^{-1}$. The functions $f_Z(\rho)$ and $f_X(\rho)$ reach asymptotic values in Eq.\,\eqref{eq:ZX boundary condition} at $\rho \gg m_Z^{-1}$ and $\rho \gg m_X^{-1}$, respectively.
The profile functions of the original gauge fields, $f_W$ and $f_A$ are given by the linear combinations of $f_Z$ and $f_X$, which converge to their asymptotic values 
in Eq.\eqref{eq: WA boundary condition} for $\rho \gg m_X^{-1}$.
For $n_\eta =0$, $F_\eta(\rho)$ remains almost flat, minimizing its contribution to the string tension.
On the other hand, $f_A(\rho)$ is slightly deformed from $f_A(\rho)=0$ for $n_\eta = 0$ due to the winding of $H_1$.
This shows that non-vanishing magnetic $A$-flux is generated along the $ZX$-string even for $n_\eta =0$, although the integrated magnetic $A$-flux is vanishing.

\begin{figure}[t]
  \centering
\includegraphics[width=0.6\textwidth]{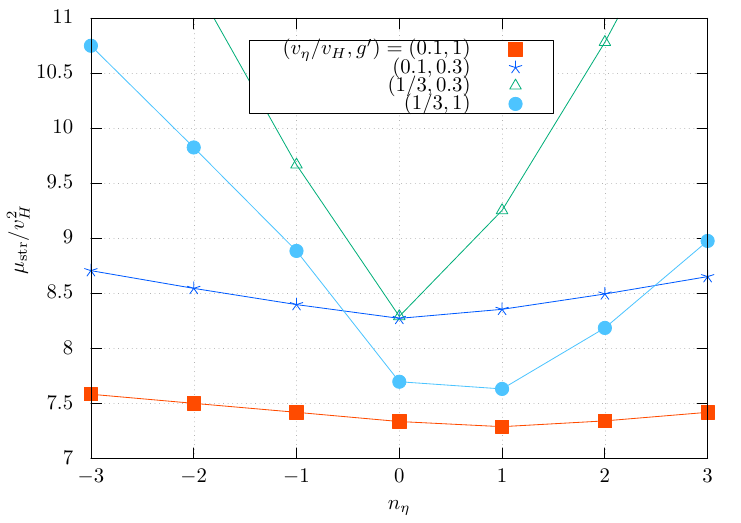}
\caption{The tension of the $ZX$-string for various winding number of $n_\eta$.
We take $\lambda_{H,\eta}$ and $g$ as in Eq.\,\eqref{eq:sample}, 
while we vary $g'$ and $v_\eta/v_H$ as indicated.
}
  \label{fig:tension}
\end{figure}
In Fig.\,\ref{fig:tension}, we show the string tension of the $ZX$-string for various winding number of $\eta$.
The figure shows that the string tension is minimized for $n_\eta = 0$,
when $g' \ll g$ and $v_\eta \ll v_H$.
This corresponds to the parameter for which 
the asymptotic value of
$f_X$ becomes minimum  (see Eq.\,\eqref{eq:asymptotic Z and X}).
Note that, in special cases such as $g' = g$ and $n_\eta / Q_\eta = 1$, the asymptotic value becomes $f_X = 0$ at $n_\eta = 1$. In such cases, the configuration with $n_\eta \neq 0$ yields a smaller tension than that with $n_\eta = 0$.

\section{Conclusions}
In this work, we have extended the analysis of metastable cosmic strings from the minimal symmetry breaking pattern $\mathrm{SU}(2)\to\mathrm{U}(1)\to \mathrm{Nothing}$ to a class of models characterized by the chain
\begin{align}
\mathrm{SU}(2)\times \mathrm{U}(1) \to \mathrm{U}(1)\times \mathrm{U}(1) \to \mathrm{U}(1)' \to \mathrm{Nothing}\ .
\end{align}
Such extended patterns emerge in various grand unified theories and cosmological models of inflation. 
In the explicit model presented in Ref.\,\cite{Chitose:2024pmz}, where inflation is driven by the $\mathrm{SU}(2)$ breaking field and monopoles are inflated away, the $\mathrm{U}(1)'$ symmetry is identified with $\mathrm{U}(1)_\mathrm{B-L}$.

A key result of our study is the demonstration that the magnetic flux carried by metastable string segments can be either confined or unconfined, depending on the sequence of symmetry breaking and string decay. If the string decays prior to the final $\mathrm{U}(1)'=\mathrm{U}(1)_\mathrm{X}$ breaking (Case A), the resulting magnetic flux remains unconfined and extends from the monopole endpoint.
In contrast, if the decay occurs after $\mathrm{U}(1)'=\mathrm{U}(1)_\mathrm{X}$ breaking (Case B), the flux becomes confined within the segment.

This distinction significantly impacts the cosmological evolution of the string network. Segments with unconfined flux efficiently lose energy through 
massless gauge boson radiation 
as well as 
interactions with $\mathrm{U}(1)_\mathrm{X}$-charged particles in the thermal plasma. Thus,
the string segments with unconfined fluxes are therefore unlikely to contribute to the gravitational wave background. On the other hand, segments with confined flux are less prone to such processes. 

Although previous studies have explored the dynamics of segments with confined or unconfined flux~\cite{Martin:1996ea,Martin:1996cp,Berezinsky:1997kd}, a detailed quantitative analysis of their evolution and corresponding gravitational wave signals remains to be performed~\cite{ChitoseIbeNedaShirai:FutureWork2025}. 
Those analyses will be particularly important
in view of upcoming 
future PTA experiments \cite{Carilli:2004nx, Janssen:2014dka, Weltman:2018zrl} and interferometer experiments \cite{LISA:2017pwj,Baker:2019nia,Blanco-Pillado:2024aca,Abac:2025saz},
 which will enable unprecedented tests of the stochastic gravitational wave background from cosmic strings. 

\section*{Acknowledgments}

This work is supported by Grant-in-Aid for Scientific Research from the Ministry of Education, Culture, Sports, Science, and Technology (MEXT), Japan,  22K03615, 24K23938 (M.I.), and by World Premier International Research Center Initiative (WPI), MEXT, Japan. 
The work of S.S. is supported by DAIKO FOUNDATION. 
This work is also supported by Grant-in-Aid for JSPS Research Fellow 
JP24KJ0832 (A.C.). 
This work is supported by FoPM, WINGS Program, the University of Tokyo (A.C.).

\appendix
\section{Mixing Angle Between \texorpdfstring{$\boldsymbol{Z}$}{Z} and 
\texorpdfstring{$\boldsymbol{X}$}{X}}
\label{sec:mixing}
In the main text, we considered only the leading‐order mixing angle in the limit $v_H \ll v_\eta$. In this appendix, we present the general result. With two VEVs, the gauge boson squared mass matrix is given by
\begin{align}
\begin{pmatrix}
M^2_{W^3W^3} & M^2_{W^3A} \\[3pt]
M^2_{W^3A} & M^2_{AA}
\end{pmatrix}
=
\begin{pmatrix}
\displaystyle \frac{g^2\,v_H^2}{2}
& \displaystyle \frac{g\,g'\,v_H^2}{2} \\[8pt]
\displaystyle \frac{g\,g'\,v_H^2}{2}
& \displaystyle \frac{g'^2\,v_H^2}{2} + 2\,g'^2Q^2\,v_\eta^2
\end{pmatrix}.
\end{align}
Hence, the squared masses of the $Z$ and $X$ bosons are
\begin{align}
m^2_{Z,X}
&= \frac{1}{4} 
\Bigl[
(g^2+g'^2)v_H^2+4g'^2Q^2v_\eta^2
\pm \sqrt{\bigl((g^2+g'^2)v_H^2+4g'^2Q^2v_\eta^2\bigr)^2
-16\,g^2g'^2Q^2\,v_H^2v_\eta^2}
\Bigr]\ ,
\end{align}
where the upper sign corresponds to the $Z$ boson and the lower sign to the $X$ boson. The mixing angle $\alpha$ is given by
\begin{align}
\tan 2\alpha
&= \frac{2\,M^2_{W^3A}}{M^2_{W^3W^3}-M^2_{AA}}
= \frac{2g\,g'\,v_H^2}
{\displaystyle (g^2-g'^2)v_H^2-4g'^2Q^2v_\eta^2}\ .
\end{align}
\section{Equations of Motion for Profile Functions}
In Sec.\,\ref{sec:numerical},
we utilized the equations of motion for $F_{H}$, $F_\eta$, $f_Z$, and $f_X$.
The equivalent set of the equations of motions for $F_{H}$, $F_\eta$, $f_W$, and $f_A$ is given by,
\begin{align}
&\frac{(\rho F_H')'}{\rho}
- \frac{1}{\rho^2}\qty(1 - \frac{f_W}{2} - \frac{f_A}{2})^2 F_H - 2\lambda_H v_H^2 (F_H^2 - 1) F_H = 0\
 ,\\
&\frac{(\rho F_\eta')'}{\rho}- \frac{1}{\rho^2}(n_\eta - Q_\eta f_A)^2 F_\eta - 2\lambda_\eta v_\eta^2 (F_\eta^2 - 1) F_\eta = 0\ ,\\
&\rho\left(\frac{f_W'}{\rho}\right)' + {g^2 v_H^2 F_H^2}\left(1 - \frac{f_W}{2} - \frac{f_A}{2}\right) = 0\ ,\\
&\rho\left(\frac{f_A'}{\rho}\right)' + {g'^2 v_H^2 F_H^2}\left(1 - \frac{f_W}{2} - \frac{f_A}{2}\right)
+ {2g'^2 Q_\eta^2 v_\eta^2 F_\eta^2}(n_\eta - Q_\eta f_A) = 0\ .
\end{align}

\bibliographystyle{apsrev4-1}
\bibliography{bibtex}

\end{document}